# Thermal Conductivity Modeling using Machine Learning Potentials: Application to Crystalline and Amorphous Silicon

Xin Qian[1], Shenyou Peng[2], Xiaobo Li,[3] Yujie Wei[2] and Ronggui Yang[1*]

[1]Department of Mechanical Engineering

University of Colorado, Boulder, CO 80309, USA

[2]The State Key Laboratory of Nonlinear Mechanics (LNM),

Institute of Mechanics, Chinese Academy of Sciences,

Beijing, 100190, PRC

[3]State Key Laboratory of Coal Combustion,

School of Energy and Power Engineering,

Huazhong University of Science and Technology,

Wuhan 430074, Hubei, China

Email: Ronggui.Yang@Colorado.Edu

ORCID: 0000-0002-3198-2014 (Xin Qian)

0000-0002-3602-6945 (Ronggui Yang)

## Abstract

First-principles based modeling on phonon dynamics and transport using density functional theory and Boltzmann transport equation has proven powerful in predicting thermal conductivity of crystalline materials, but it remains unfeasible for modeling complex crystals and disordered solids due to the prohibitive computational cost to capture the disordered structure, especially when the quasiparticle "phonon" model breaks down. Recently, machine-learning regression algorithms show great promises for building high-accuracy potential fields for atomistic modeling with length and time scales far beyond those achievable by first-principles calculations. In this work, using both crystalline and amorphous silicon as examples, we develop machine learning based potential fields for predicting thermal conductivity. The machine learning based interatomic potential is derived from density functional theory calculations by stochastically sampling the potential energy surface in the configurational space. The thermal conductivities of both amorphous and crystalline silicon are then calculated using equilibrium molecular dynamics, which agree well with experimental measurements. This work documents the procedure for training the machine-learning based potentials for modeling thermal conductivity, and demonstrates that machine-learning based potential can be a promising tool for modeling thermal conductivity of both crystalline and amorphous materials with strong disorder.

In the past decade, first-principles based calculations have become a powerful tool for predicting thermal conductivity of a wide range of bulk[1-6] and low-dimensional crystals.[7-18] Despite the successful application in modeling the phonon properties and thermal conductivity of simple crystals, first-principles calculation for complex crystals and disordered materials remains challenging, since the computational cost increases dramatically with the lowering of crystalline symmetry and the increasing size of unit cells.[19, 20] It also becomes questionable to apply the quasiparticle "phonon" picture assumed by the Boltzmann transport theory to complex crystals and disordered materials, since considerable amount of vibrational modes in these low-symmetry systems become diffusive or localized.[21, 22] For modeling disordered systems, molecular dynamics (MD) simulations become a great choice since MD can easily incorporate detailed atomic structures including defects and local strains. However, MD has limited fidelity and accuracy, due to the lack of accurate interatomic potentials. Improving the accuracy of empirical interatomic potential is difficult, because the *ab-initio* potential energy surface (PES) can hardly be fitted by simple functional forms that are artificially assigned based on the pre-knowledge of the interatomic bonding.[19, 20, 23, 24] Using "rigid" or "definite" functional forms also severely limits the transferability among different atomic structures and material phases, since it usually requires re-formulating the fitting functionals.

Recently, machine learning potential (MLP) emerges as a promising tool for bridging the gap between the first-principles calculations and MD simulations for modeling thermal transport. Since MLP does not artificially assign functional forms, it does not suffer from the limited accuracy as empirical potential does while intrinsically incorporates anharmonic effects.[25] In the past five years, MLP has been successfully developed and used to model the structural, thermodynamic and mechanical properties of some simple crystals such as Si, Ge and GaN[26-28] and amorphous materials.[29, 30] However, implementations of MLP for studying thermal transport has been rare, and limited to simple crystals with relative weak disorder such as vacancies and alloys.[31, 32] In this work, we develop MLP for modeling thermal conductivity of both crystalline and amorphous materials, using silicon as an example. Although there exist quite a few methods of constructing MLP such as artificial neural network,[26] supporting vector regression,[33] and spectral neighbor analysis potential (SNAP),[34] the Gaussian approximation potential (GAP)[27, 35] is chosen in this work since GAP has the highest accuracy of predicting interatomic forces in comparison with other MLP methods.[36] Furthermore, the training of GAP models based on Gaussian process regression[37] only involves linear algebra without nonlinear optimizations.

First, we briefly describe here the training of the GAP models for c-Si and a-Si. To construct the training database, we use a stochastic method to generate random uncorrelated snapshots to sample the *ab-initio* PES. After GAP models are developed, thermal conductivity of both crystalline silicon (c-Si) and amorphous silicon (a-Si) are calculated using the equilibrium molecular dynamics (EMD) simulations. Figure 1a shows the training strategy for building the GAP model for c-Si. Since our goal is to model thermal conductivity, the GAP model should be able to fit and interpolate the PES around the equilibrium configuration accessible by thermal vibrations. Compared with *ab-initio* molecular dynamics which sample a trajectory in the configurational space, a more efficient method is to stochastically generate uncorrelated snapshots with random displacements:[38]

$$\boldsymbol{u}_i = \sum_s \boldsymbol{e}_{is} \langle A_{is}\rangle \sqrt{-2\ln\zeta_1} \sin(2\pi\zeta_2) \tag{1}$$

where $\boldsymbol{u}_i$ is the displacement of atom $i$ from equilibrium position, $\boldsymbol{e}_{is}$ and $\langle A_{is}\rangle$ are the eigenvector and average amplitude of atom $i$ participating in the vibration of normal mode $s$, $\zeta_1$ and $\zeta_2$ are two random numbers uniformly distributed in the interval of (0,1). The amplitude $\langle A_{is}(T) \rangle$ of normal mode $s$ can be written as:[38]

$$\langle A_{is} \rangle = \sqrt{\frac{\hbar\left(n_s + \frac{1}{2}\right)}{m_i \omega_s}} \tag{2}$$

where $m_i$ is the mass of atom $i$, $n_s$ is the Bose-Einstein distribution $n_s = \left[\exp\left(\frac{\hbar \omega_s}{k_B T}\right) - 1\right]^{-1}$ at temperature $T$ and frequency $\omega_s$.

Clearly, generating displacement snapshots using Eqs. (1-2) requires the knowledge of force constants to obtain both normal mode frequency $\omega_s$ and eigenvectors $\boldsymbol{e}_{is}$. For c-Si, $\omega_s$ and $\langle A_{is} \rangle$ can be easily obtained from harmonic lattice dynamics calculations. Phonopy[39] package is used to generate supercells of c-Si containing 5×5×5 primitive cells with finite displacements, then density functional theory (DFT) calculations are performed to obtain the interatomic forces using the projected augumented wave (PAW) method implemented in the VASP package.[40, 41] Exchange-correlation energy is treated with the Perdew–Burke-Ernzerhof (PBE) functionals,[42] and the plane-wave cutoff energy is set to 350 eV, more than 40% larger than the maximum plane-wave energy recommended by the pseudopotential for Si.[43, 44] After the vibrational frequencies and eigenvectors are obtained by solving the dynamical equation. The displacement amplitudes are calculated at 300 K and 600 K, with 100 snapshots at each temperature. In total, 50000 local chemical environments (i.e. each atom with its neighbors in each snapshot) are sampled in the database. Self-consistent field (SCF) calculations are then performed using the VASP package for each snapshot to obtain the energies and forces, which are then used as target observables to be fitted in the training dataset. In the training process, the energies and forces are expressed as a linear combination of a set of kernel functions specified in the SOAP descriptor,[45] and the associated linear coefficients are obtained through the sparsified Gaussian process regression formalism[46] where the regression details can be found in ref. [47]. Hyper-parameters used for training including cutoff radius ($r_{cut}$) of neighboring atoms, smoothing width of atom density distribution ($\sigma_a$) and others are listed in Table 1 while the meaning of each parameter can be found in our previous work[25] and references [29, 31].

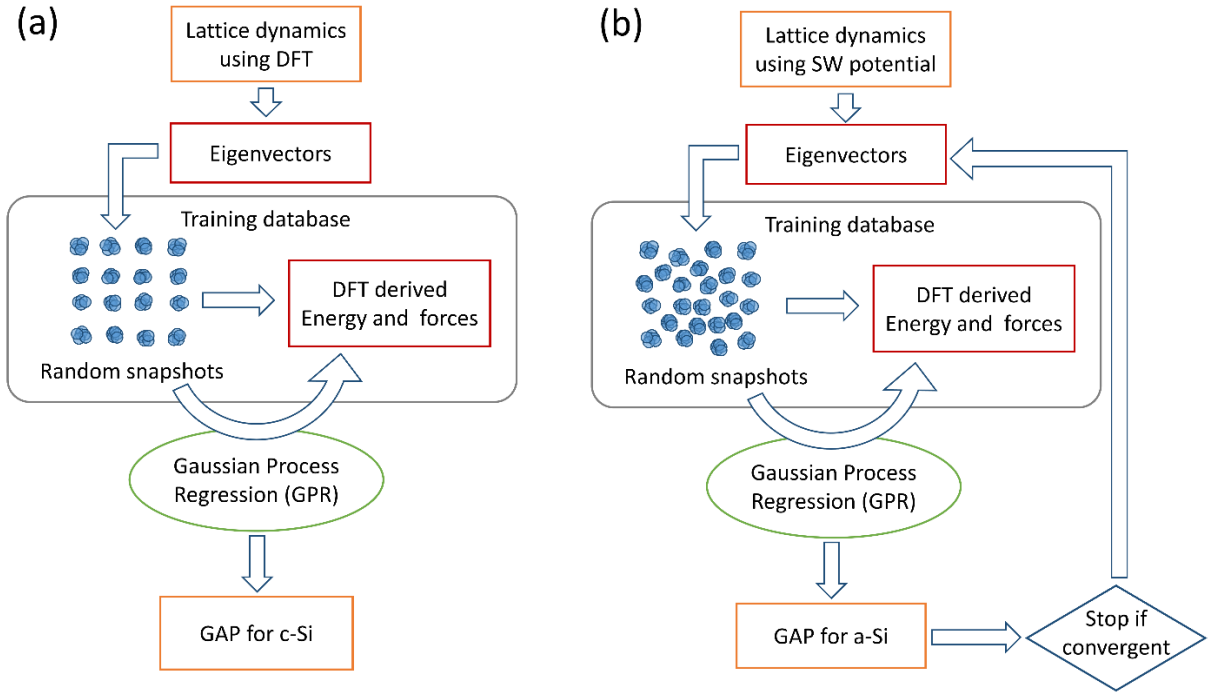

*Figure 1. (a) Training strategy for c-Si: lattice dynamics using are performed with finite-displacement method using DFT. to obtain eigenvectors which is then used to generate snapshots with random displacements. DFT calculations are performed to obtain energies and forces corresponding to these snapshots. The energies and forces are then used as the training database to obtain GAP model for c-Si. (b) Training strategy for a-Si. In the first iteration of training, eigenvectors necessary to generate random snapshots are obtained from the empirical SW potential. DFT calculations are performed to obtain energies and forces corresponding to these snapshots. The energies and forces are then used as the training database to the first generation of GAP. A new set of eigenvectors are derived from the GAP model, which are used to train the next generation of GAP model. Such process is repeated until the energy change is smaller than $2\times10^{-3}$ eV/atom.*

Table 1. Hyper-parameters for GAP with SOAP kernels.

| | |
|---|---|
| $r_{cut}$ | 4.5 Å |
| $d$ | 0.5 Å |
| $\sigma_v$ for energy | 0.0001 eV/atom |
| $\sigma_v$ for forces | 0.001 eV/Å |
| $\sigma_w$ | 1.0 eV |
| $\sigma_a$ | 0.5 Å |
| $\zeta$ | 4 |
| $n_{max}$ | 12 |
| $l_{max}$ | 12 |

However, constructing the training database for a-Si is not as simple. First, it is nontrivial to obtain a relaxed amorphous network with atoms in equilibrium positions from DFT. Since a reasonable initial structure is important for the convergence of energy and forces when relaxing the atomic structures using DFT calculations, the classical MD simulation was performed first using Stillinger-Weber (SW) potential[48] to generate the initial structure of a-Si using a melt-quench method.[30] A c-Si simulation cell containing 216 atoms ($3 \times 3 \times 3$ conventional cells) is first thermalized to 3000 K for 500 ps using Nose-Hoover reservoir (NPT ensemble) for the melting process. The temperature of Nose-Hoover reservoir is then decreased to 1 K with a rate of 10 K/ps, and is kept at the final temperature to quench the system for another 2 ns. The final structure obtained from classical MD is then used as an initial guess for the amorphous Si network for performing geometry relaxation in DFT calculations. Although larger a-Si simulation cells can be obtained using this melt-quench method, the number of atoms is limited to 64 to 216 atoms accessible by DFT calculations due to the computational cost.[49] The simulation cell obtained using SW potential is then relaxed using the conjugated gradient (CG) algorithm implemented in VASP package,[41] until the atomic forces becomes smaller than $10^{-6}$ eV/Å. The other challenge is the high computational cost to perform first-principles lattice dynamics on an amorphous network containing hundreds of atoms. To mitigate this challenge, we use the training strategy for a-Si as shown Figure 1b. Instead of generating random snapshots directly from DFT calculations, a set of trial eigenvectors is derived using the empirical SW potential. Since the optimized structure obtained from DFT calculations is not the same as the equilibrium structure obtained with the SW potential, there are soft vibrational modes with imaginary frequencies when performing lattice dynamics using SW potential. These soft modes are excluded when using Eq. (1) to sum the displacement over all modes for the first generation of random snapshots. After obtaining the trial snapshots, SCF calculations of the energy and forces were performed for each snapshot. The obtained forces and energies of the snapshots are recorded in the database for training the first generation of GAP model. Note that the first generation of the generated random displacements do not correspond to the equilibrium population of the phonon modes. To minimize the possible error induced by the unphysical displacements, we adopted an iterative training process similar to the method used by Hellman *et al.* for developing temperature dependent effective potential (TDEP) method.[38] The first generation of GAP model is used to perform lattice dynamics again to generate a new set of snapshots for training the next generation of GAP model. Such process is repeated until the change of the total energy is smaller than $2\times10^{-3}$ eV/atom, and the soft modes disappear to ensure that the trained structure is dynamically stable using the GAP model. In this work, 50 snapshots are generated for training each generation of GAP model, and convergence of atomic energy is achieved in the third iteration of training.

After the GAP models are trained, it is necessary to evaluate not only the accuracy of GAP models for reproducing the *ab-initio* energies and forces in the training database, but also the accuracy in predicting energies and forces for snapshots that are not in the training database. The root-mean-squared error (RMSE) of the energies and forces of GAP models are calculated by comparing with the data in the training databases. As shown in Figure 2a-b, the RMSE of GAP reproducing the energy and interatomic forces in c-Si are 0.00057 eV/atom and 0.0215 eV/Å, respectively. Compared with the RSME of interatomic forces (0.29 eV/ Å) using empirical SW potential, the RMSE of forces using GAP model is one order of magnitude smaller. Clearly shown in Figure 2b, the SW potential has a steeper correlation to DFT forces compared with GAP models, which means that SW systematically overestimated the interatomic forces. Compared with our previous work using *ab-initio* molecular dynamics (AIMD) simulations for sampling the PES for crystalline Zr,[25] similar level of regression accuracy is achieved while the required number of snapshots is one order of magnitude smaller. AIMD snapshots sample a trajectory on the *ab-initio* PES, which are inter-correlated. The stochastically generated snapshots are independent of each other which results in a more effective sampling in the configurational space. To evaluate the accuracy in predicting forces of snapshots outside the training database, another set of snapshots was generated as the testing dataset and the corresponding forces and energies are calculated, as shown in Figure 2a-b. The RMSE of energy and interatomic forces evaluated based on testing dataset are 0.00058 eV/atom and 0.0217 eV/Å, respectively, which are very close to the accuracy evaluated based on the training dataset. As shown in Figure 2c, phonon dispersion for c-Si using GAP is in excellent agreement with DFT calculation based on Purdue-Burke-Ernzerhof (PBE) functional, indicating that the GAP model can accurately reproduce the harmonic force constants.

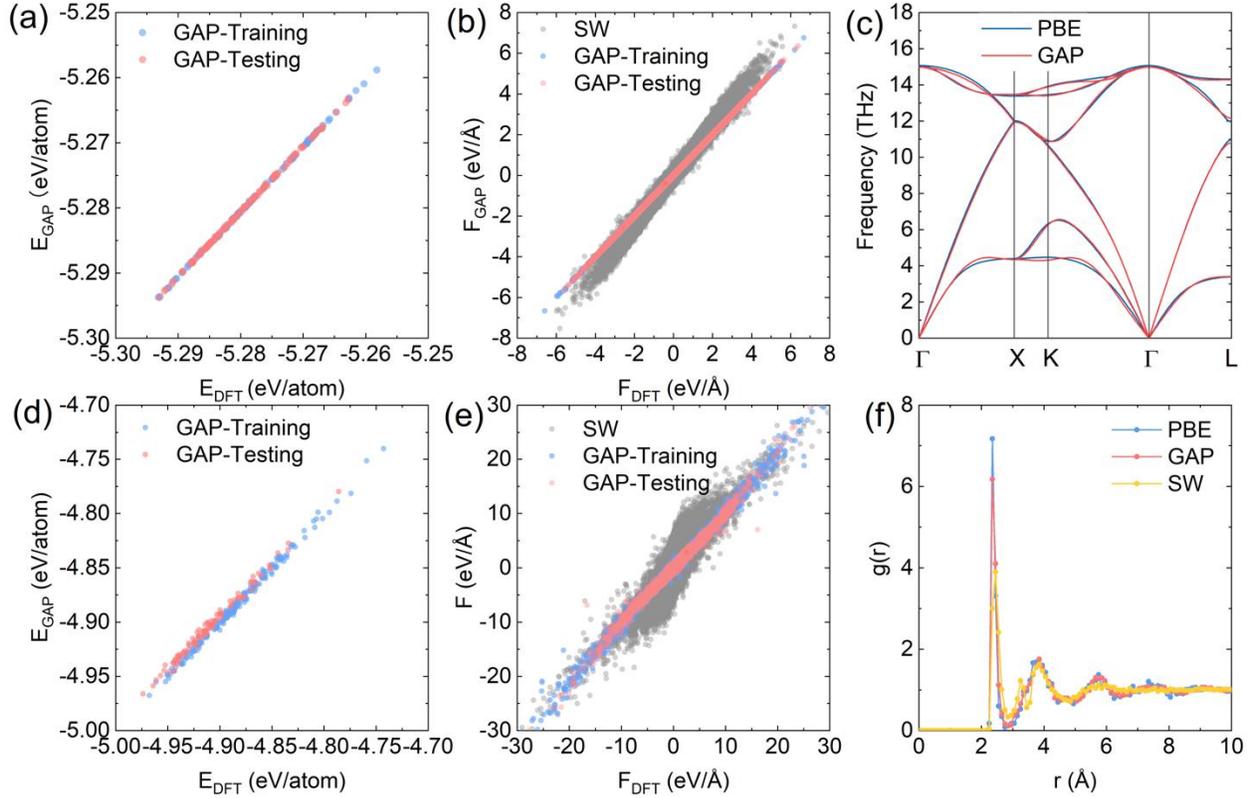

*Figure 2. Comparisons of (a) energy and (b) forces computed from DFT and GAP, and (c) phonon dispersion in c-Si. Comparisons of (d) energy and (e) forces computed from DFT using PBE functional and GAP for a-Si. (f) Radial distribution function of equilibrium a-Si structure predicted by DFT using PBE functional, GAP and the empirical SW potential.*

Figure 2d-e shows regression and prediction accuracy of energy and forces of a-Si. In the training database, the RMSE for reproducing energy and forces (0.034 eV/atom and 0.34 eV/Å, respectively) is one order of magnitude larger compared with the c-Si, due to the much more complicated atomic structure and local atomic environments.[29] The accuracy in predicting energy and forces becomes lower when using the training dataset. The RMSE of energy and forces are 0.066 eV/atom and 0.54 eV/Å evaluated using the testing database, respectively. Similar to the case of c-Si, the accuracy of calculating interatomic forces using GAP models still outperform the empirical SW potential by an order of magnitude, whose RMSE for the forces is as big as 1.5 eV/

Å, similar to the case in c-Si. To further assess whether GAP could accurately capture the structural features of a-Si network, the radial distribution function (RDF) $g(r)$ of the equilibrium a-Si structures obtained from GAP and PBE are calculated and compared, as shown in Figure 2f. It is observed that GAP can reasonably reproduce the radial distribution function compared with PBE functionals, while SW potential falsely predicts a peak of RDF near 3 Å.

After the training process, GAP models are developed to predict interatomic forces need for thermal conductivity calculations. Equilibrium molecular dynamics (EMD) are performed to obtain thermal conductivity using the LAMMPS package.[50] First, the isothermal-isobaric ensemble (NPT) are used to thermalize the simulation cells for 400 ps with a time step of 0.5 fs for both c-Si and a-Si. The simulations are then switched to microcanonical (NVE) ensemble for thermal conductivity calculation. In GAP models, heat flux $\boldsymbol{J}$ is expressed as:

$$\boldsymbol{J} = \frac{1}{V} \sum_i (E_i \boldsymbol{v}_i - \boldsymbol{S}_i \cdot \boldsymbol{v}_i) \qquad (3)$$

where $V$ is the volume of the simulation cell, $E_i$ and $\boldsymbol{v}_i$ are the energy and velocity of atom $i$, and the atomic virial stress tensor $\boldsymbol{S}_i$ is written as the outer product of relative position $\boldsymbol{r}_j - \boldsymbol{r}_i$ and local potential derivative with respect to the neighboring atom $\frac{\partial E_i}{\partial \boldsymbol{r}_j}$:

$$\boldsymbol{S}_i = \sum_j (\boldsymbol{r}_j - \boldsymbol{r}_i) \otimes \frac{\partial E_i}{\partial \boldsymbol{r}_j} \qquad (4)$$

Thermal conductivity is then calculated using the Green-Kubo formula:

$$k = \frac{V}{3k_B T^2} \int \langle \boldsymbol{J}(0) \cdot \boldsymbol{J}(t) \rangle dt \qquad (5)$$

To perform the Green-Kubo integration, heat autocorrelation function $\langle \boldsymbol{J}(0) \cdot \boldsymbol{J}(t) \rangle$ is sampled every 5 fs, and integrated up to 200 ps for c-Si and 40 ps for a-Si until the thermal conductivity values stopped increasing with the increase of the correlation time, as shown in Figure 3a-b. To suppress the uncertainty, ten individual simulations with different initial velocity distributions are performed to average the heat autocorrelation function. At room temperature, the thermal conductivity values of c-Si and a-Si are found to be 121 W/mK and 1.4 W/mK, respectively. In Figure 3c, the thermal conductivity of c-Si obtained from EMD is compared with the values calculated by iteratively solving Boltzmann transport equation, using the harmonic and third-order force constants obtained from PBE functionals (145 W/mK) and the trained GAP model (137 W/mK). ShengBTE package[43] is used to iteratively solve the BTE with a 13×13×13 $q$-mesh to sample the reciprocal $q$ space. The slightly lower thermal conductivity obtained by solving BTE from GAP could be attributed to the error in predicting the interatomic forces. Even with the same GAP model, the thermal conductivity predicted by EMD simulation is 12% lower than that by solving BTE. There could be two reasons leading to the lower thermal conductivity from EMD simulations: 1). the classical statistics in MD could lead to overestimated scattering rates,[51] and 2) MD simulations with GAP naturally includes higher order anharmonicities, while BTE approach truncates the anharmonic force constants to the third order. At a high temperature of 500 K, the difference between thermal conductivity predicted by EMD (61 W/mK) and BTE (77 W/mK) increased to ~20%, indicating that GAP tends to overestimate the higher-order anharmonicity. Such underestimated thermal conductivity by EMD using machine learning based interatomic potential is also observed in transition metal dichalcogenide alloys.[31] Compared with the empirical SW potential,[52] GAP model still shows much higher accuracy in predicting thermal conductivity of c-Si. Figure 3d shows thermal conductivity of a-Si obtained from EMD using the GAP for a-Si. The predicted thermal conductivity is ~1.4 W/mK at room temperature, within the range of measurement values of a-Si (1 ~ 2 W/mK). [53-55] Considering the fact that experimentally prepared a-Si usually contains different concentrations of hydrogen which reduces the phonon localization and leads to higher thermal conductivity,[56] the thermal conductivity obtained in this could serve as an estimation for non-hydrogenated a-Si. However, the GAP model predicts a lower thermal conductivity than using the empirical potentials, [57-59] probably due to the fact that the empirical potentials predicts higher bonding stiffness compared with DFT calculation using PBE functionals, consistent with the trend we observed in c-Si.

In summary, we have developed GAP models with regression accuracy of 0.02 eV/Å and 0.3 eV/Å for interatomic forces in crystalline and amorphous silicon, respectively, showing one-order-of-magnitude improvement in both energy and forces compared with the empirical SW potential. This work also shows that the training database can be efficiently constructed using stochastically generated snapshots with much smaller amount of DFT calculations than AIMD, while achieving similar accuracy of fitting the *ab-initio* potential energy surface. Thermal conductivity of c-Si and a-Si at room temperature is calculated to be 121 W/mK and 1.4 W/mK respectively using EMD, agreeing reasonably well with experiments and first-principles calculations. This work shows that GAP can be a promising tool for modeling thermal conductivity of both crystalline and amorphous materials with strong disorder.

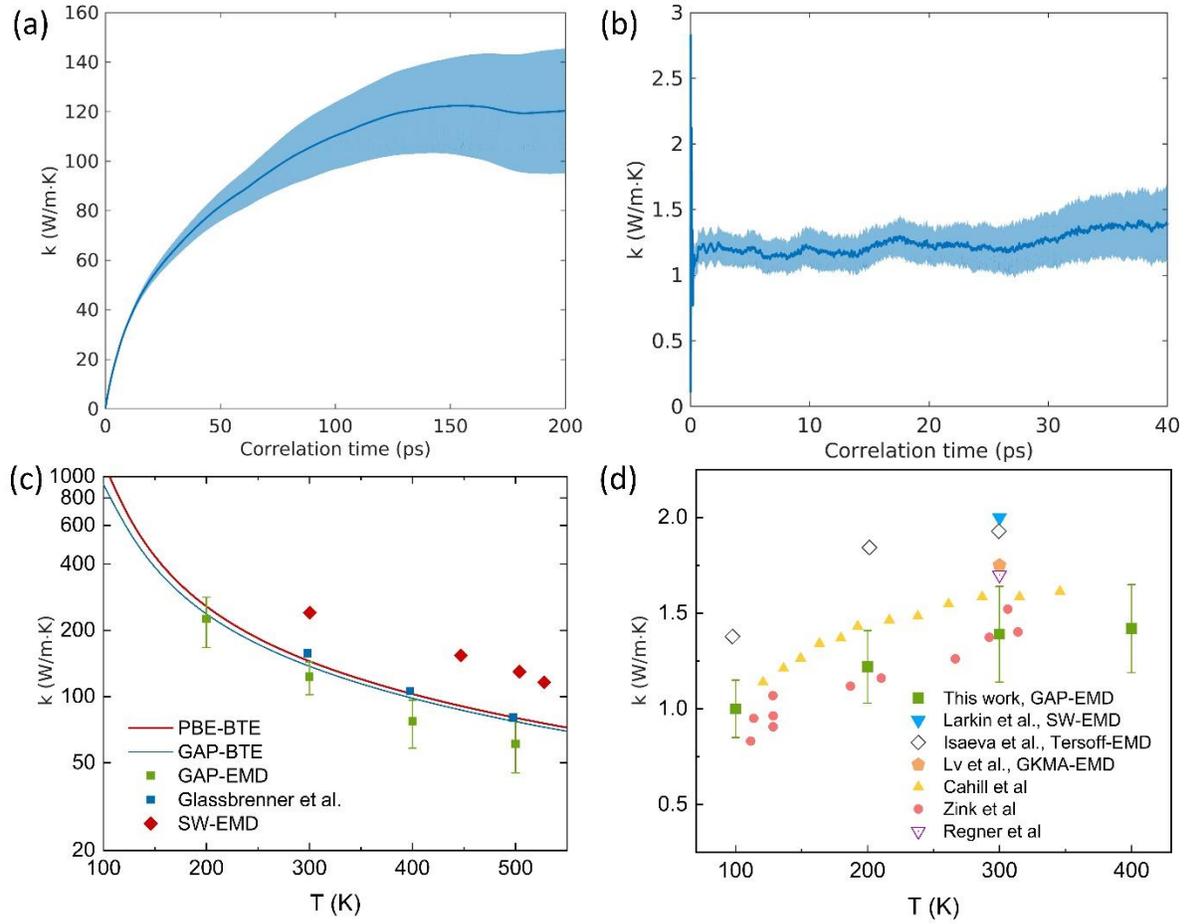

*Figure 3. (a-b) Thermal conductivity calculated using Green-Kubo method of (a) c-Si and (b) a-Si as a function of correlation time. The shaded area shows the standard deviation among 10 independent simulations with different initial velocity distributions (c) Thermal conductivity of crystalline silicon derived from GAP model using EMD and BTE, compared with experiments by Glassbrenner et al,[60] and EMD simulation using SW potential by Volz et al.[52] (d) Thermal conductivity of amorphous silicon calculated using EMD with the GAP model developed in this work, compared with EMD results by Larkin et al.,[57] Isaeva et al.[58] and Lv et al.,[59] and experimental measurements by Zink et. al[53]., Regner et al.,[54] and Cahill et. al.[55]*

**Acknowledgement:** This work is supported by NSF (Grant No. 1512776). Density functional theory calculations and training of the GAP models are performed using the Summit supercomputer, which is supported by NSF (awards ACI-1532235 and ACI-1532236), University of Colorado Boulder and Colorado State University. Molecular dynamics simulations are performed using the Yuan supercomputer supported by Supercomputing Center of Chinese Academy of Sciences.